\documentclass[twoside,twocolumn,9pt]{article}
\usepackage{extsizes}
\usepackage[super,sort&compress,comma]{natbib} 
\usepackage[version=3]{mhchem}
\usepackage[left=1.5cm, right=1.5cm, top=1.785cm, bottom=2.0cm]{geometry}
\usepackage{balance}
\usepackage{mathptmx}
\usepackage{sectsty}
\usepackage{graphicx} 
\usepackage{lastpage}
\usepackage[format=plain,justification=justified,singlelinecheck=false,font={stretch=1.125,small,sf},labelfont=bf,labelsep=space]{caption}
\usepackage{float}
\usepackage{fancyhdr}
\usepackage{fnpos}
\usepackage{gensymb}
\usepackage[english]{babel}
\addto{\captionsenglish}{%
  
}
\usepackage{subfig}
\usepackage{array}
\usepackage{droidsans}
\usepackage{charter}
\usepackage[T1]{fontenc}
\usepackage[usenames,dvipsnames]{xcolor}
\usepackage{setspace}
\usepackage[compact]{titlesec}
\usepackage{hyperref}

\usepackage{epstopdf}

\definecolor{cream}{RGB}{222,217,201}

\begin{document}

\pagestyle{fancy}
\thispagestyle{plain}
\fancypagestyle{plain}{
\renewcommand{\headrulewidth}{0pt}
}

\makeFNbottom
\makeatletter
\renewcommand\LARGE{\@setfontsize\LARGE{15pt}{17}}
\renewcommand\Large{\@setfontsize\Large{12pt}{14}}
\renewcommand\large{\@setfontsize\large{10pt}{12}}
\renewcommand\footnotesize{\@setfontsize\footnotesize{7pt}{10}}
\makeatother

\renewcommand{\thefootnote}{\fnsymbol{footnote}}
\renewcommand\footnoterule{\vspace*{1pt}%
\color{cream}\hrule width 3.5in height 0.4pt \color{black}\vspace*{5pt}} 
\setcounter{secnumdepth}{5}

\makeatletter 
\renewcommand\@biblabel[1]{#1}            
\renewcommand\@makefntext[1]%
{\noindent\makebox[0pt][r]{\@thefnmark\,}#1}
\makeatother 
\renewcommand{\figurename}{\small{Fig.}~}
\sectionfont{\sffamily\Large}
\subsectionfont{\normalsize}
\subsubsectionfont{\bf}
\setstretch{1.125} 
\setlength{\skip\footins}{0.8cm}
\setlength{\footnotesep}{0.25cm}
\setlength{\jot}{10pt}
\titlespacing*{\section}{0pt}{4pt}{4pt}
\titlespacing*{\subsection}{0pt}{15pt}{1pt}

\fancyfoot{}
\fancyfoot[LO,RE]{\vspace{-7.1pt}\includegraphics[height=9pt]{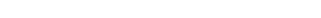}}
\fancyfoot[CO]{\vspace{-7.1pt}\hspace{13.2cm}\includegraphics{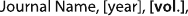}}
\fancyfoot[CE]{\vspace{-7.2pt}\hspace{-14.2cm}\includegraphics{head_foot/RF}}
\fancyfoot[RO]{\footnotesize{\sffamily{1--\pageref{LastPage} ~\textbar  \hspace{2pt}\thepage}}}
\fancyfoot[LE]{\footnotesize{\sffamily{\thepage~\textbar\hspace{3.45cm} 1--\pageref{LastPage}}}}
\fancyhead{}
\renewcommand{\headrulewidth}{0pt} 
\renewcommand{\footrulewidth}{0pt}
\setlength{\arrayrulewidth}{1pt}
\setlength{\columnsep}{6.5mm}
\setlength\bibsep{1pt}

\makeatletter 
\newlength{\figrulesep} 
\setlength{\figrulesep}{0.5\textfloatsep} 

\newcommand{\topfigrule}{\vspace*{-1pt}%
\noindent{\color{cream}\rule[-\figrulesep]{\columnwidth}{1.5pt}} }

\newcommand{\botfigrule}{\vspace*{-2pt}%
\noindent{\color{cream}\rule[\figrulesep]{\columnwidth}{1.5pt}} }

\newcommand{\dblfigrule}{\vspace*{-1pt}%
\noindent{\color{cream}\rule[-\figrulesep]{\textwidth}{1.5pt}} }

\makeatother

\twocolumn[
  \begin{@twocolumnfalse}
{\includegraphics[height=30pt]{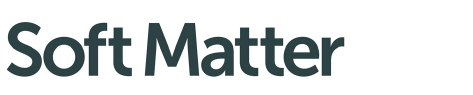}\hfill\raisebox{0pt}[0pt][0pt]{\includegraphics[height=55pt]{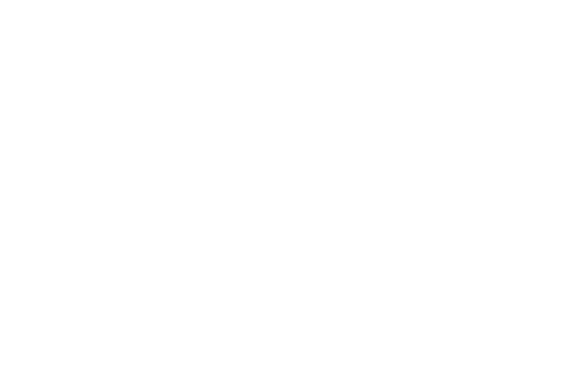}}\\[1ex]
\includegraphics[width=18.5cm]{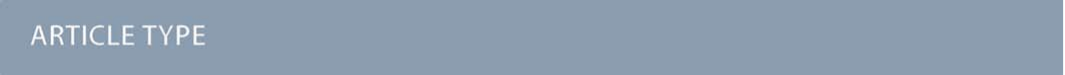}}\par
\vspace{1em}
\sffamily
\begin{tabular}{m{4.5cm} p{13.5cm} }

\includegraphics{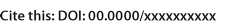} & \noindent\LARGE{\textbf{Mesoscale heterogeneity in protein hydrogels induced by dynamic unfolding and post-gelation rearrangements}} \\
\vspace{0.3cm} & \vspace{0.3cm} \\

 & \noindent\large{Victoria Byelova,\textit{$^{a}$} Lorna Dougan,\textit{$^{a}$} and David Head$^{\ast}$\textit{$^{b}$}} \\

\includegraphics{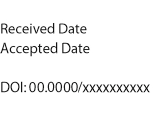} & \noindent\normalsize{The rationalization of protein hydrogel design is essential in creating bespoke and functional materials for a range of medical and healthcare applications, such as tissue engineering, wound healing and bio-sensors. The dynamic process of protein unfolding can significantly influence the resultant gel structure and behaviour - conformational changes in individual protein building blocks affect the bulk gel, whilst changes to bulk conditions influence how individual proteins move and respond. In order to characterize and control protein hydrogels, a multi-lengthscale picture must be built so that this complex interplay can be understood. We develop a coarse-grained computational model to investigate fundamental aspects of dynamic unfolding within chemically crosslinked protein hydrogels, focussing on the mesoscale structure during and after gelation. In simplifying protein unfolding to a single dynamic step, we are able to create a heterogeneous network formed of coarse-grained proteins. We observe evolving hydrogels consisting of regions of high-density protein clusters with a low degree of unfolding due to local crowding effects, interconnected by low-density stranded regions with a higher concentration of unfolded proteins. Systematically varying the barrier height for single protein unfolding demonstrates a weak influence on the mesoscale structure as characterised by fractal dimension and correlation length. Conversely, the measured changes in these parameters are greater after the gel experiences sustained conformational relaxation. This work demonstrates the effectiveness of coarse-grained modelling in capturing the heterogeneous mesoscale network of crosslinked unfoldable proteins. A greater predictive understanding of the structure and behaviour of protein hydrogels is gained through this model and will accelerate the design of novel biomaterials.} \\

\end{tabular}

 \end{@twocolumnfalse} \vspace{0.6cm}

  ]

\renewcommand*\rmdefault{bch}\normalfont\upshape
\rmfamily
\section*{}
\vspace{-1cm}


\footnotetext{\textit{$^{a}$~School of Physics and Astronomy, University of Leeds, Leeds, United Kingdom}}
\footnotetext{\textit{$^{b}$~School of Computer Science, University of Leeds, Leeds, United Kingdom }}




\section*{Introduction}
Protein hydrogels are versatile materials with controllable multi-lengthscale behaviour and increasing capacity to be rationally designed\cite{kong2014afm, chen2021b, mout2024pnasu, narayan2025ab}. By selecting a protein building block and chemically crosslinking in an aqueous environment, an irreversibly crosslinked protein network is produced. Protein networks are naturally abundant in many biological systems and often rely on protein unfolding for functionality, whether it is for mechanotransduction\cite{kubow2015nc}, hierarchical structure control\cite{booth1997n} or to be stimuli-responsive\cite{rief1997s}. By understanding living systems it has been possible to create synthetic protein hydrogels for applications such as  tissue engineering \cite{leach2005b, gonen-wadmany2007b, meganathan2022ba}, wound healing\cite{ouyang2022bm, li2024cej, lu2025nr} and wearable sensors\cite{yang2022ma, xu2023ijobm, li2024icap}. Protein hydrogels also are frequently studied and manufactured to enhance nutrition and mouthfeel in foods \cite{li2024fh, amin2024fh, sun2026ijobm}. There is evidence to suggest that protein unfolding during gel network formation changes resultant gel structure and mechanics\cite{hughes2021an, hughes2025sm, hughes2025jocais, kong2014afm, shmilovich2018prl, khoury2019nc, nowitzke2024an, lebon1999m, gimel1994m} - certain proteins are selected as building blocks for this purpose. This building block may be a designed synthetic protein \cite{fang2013nc} or a naturally-occurring protein with an increased probability of unfolding due to the addition of a reducing agent.\cite{sikora2011p}. By creating a pre-gel solution with proteins that can unfold, a complex interplay arises between individual protein behaviour, bulk gel behaviour and how these two affect one another once network formation begins. However the specific relation between the two is difficult to experimentally verify, especially with complex multi-step protein unfolding pathways\cite{zocchi1997bcbhf, rief1997s, brockwell2005bj, dietz2004pnasu, fernandez2004s, gebhardt2010pnasu, kellermayer1997s, tskhovrebova1997n}. It is possible to characterise a single-protein mechanical unfolding landscape and use this to inform how the bulk gel may behave\cite{wu2018nc} but the challenge lies in generalizing this relation for a broad range of protein hydrogel systems.\\\\
Whilst \textit{in-vitro} synthesis and analysis of protein hydrogels can provide insight into bulk gel properties, a greater difficulty lies in creating a multi-lengthscale representation of a system. Gel design and characterisation is frequently informed by single protein building block design and bulk analysis of the chemically crosslinked gel\cite{huerta-lopez2024sa} but the mesoscale in-between is less commonly characterized - there is difficulty in acquiring precise information about cluster-cluster interactions and localising where certain protein behaviours occur in a network. Mesoscale interactions and features of protein gels during network formation are highly influential on the resultant gel properties\cite{poon1997} - however, how individual protein-protein interactions collectively contribute to corresponding bulk network behaviour is poorly understood. As such it is appropriate to employ computational simulations of protein networks to gain a controlled and precise view into network properties. Single protein unfolding landscapes have been investigated with varying degrees of coarse-graining to understand unfolding transitions\cite{minin2017jacs, zhmurov2012jacs, hirschmann2023jcp}. Regarding network simulations, a variety of methods have previously been used with focus on different protein features such as the use of patchy particles for colloidal network formation \cite{zhang2004nl, hanson2020m}, hindered rotation for the purpose of stiffer gels\cite{nguyen2020sm} and a multidomain-protein system that evaulated the importance of directional mechanical unfolding\cite{nowitzke2024an}. This approach created small networks of 8-bead polymers to replicate the mechanical response of polyprotein L, and found that initial rearrangements and unfolding of polyproteins led to enhanced unfolding upon rearrangement of proteins to be oriented to the force vector. Another model studied the role of diffusion- or  reaction-limited aggregation of globular, folded protein gels confined to a lattice, affirming that reaction rate influences network formation and structure\cite{jungblut2019pccp, cook2023sm} - such an approach that chooses to remove additional detail to focus on one key parameter in greater depth is referred to here as `coarse-grained.' Though no protein-specific parameters or explicit (hydro-)dynamics were included, it was possible to extract information about factors influencing protein network formation such as volume fraction and reaction probability, highlighting the benefits of using a coarse-grained approach to model large three-dimensional systems. \\\\
As a model to investigate the effects of unfolding in networks, prior experimental work\cite{hughes2021an} compared gels formed using mechanically robust, globular bovine serum albumin (BSA) and gels formed with BSA that had been solvated and had an increased probability of unfolding. It was found that gels containing a population of folded and unfolded proteins produced more heterogeneous gels with enhanced storage and loss moduli, as well as improved energy dissipation. Coarse-grained simulations were also performed to emulate these two gel types - systems replicating gels with an increased probability of unfolding were simulated as a mixture of folded and unfolded proteins without transitions between the two forms. The model used an open-source software package BioNet, which represents folded, globular proteins as sticky spheres\cite{hanson2019sm}. Whilst \textit{in-vitro} analysis found that solvated BSA gels produced more heterogeneous structures than in non-solvated gels, simulations saw that mixed gels were less heterogeneous than purely folded BSA gels, creating a disparity between \textit{in-silico} and \textit{in-vitro} techniques. As the simulated gels were mixed state systems, the study inferred that the dynamic process of unfolding is required to produce the difference in structural heterogeneity of networks that is seen during \textit{in-vitro} experiments.\\\\
Other coarse-grained computational models of protein hydrogels have focused on a variety of parameters contributing to network formation and behaviour. The multidomain system of protein L focussed on network rearrangements propagated by directional mechanical unfolding \cite{nowitzke2024an}, whilst the sticky sphere model used to build a large colloidal network has been used to understand the contribution of volume fraction, number of binding sites and mixed systems to network structure \cite{hanson2020m, hanson2021b}. Though colloidal models for hydrogel networks provide an excellent basis for understanding aggregation and relaxation\cite{poon1997, colombo2014jr, shih1990pra, wu2001l, hagiwara1998}, these lack the contribution of single protein unfolding to network growth and rearrangement. A large-scale network is also required so that it is possible to localise regions of high probability of unfolding, and how this influences the protein network more broadly. With these previous experiments and simulations in mind, we present a coarse-grained model for a protein building block that uses a single-step unfolding process. This singular step represents the beginnings of single protein unfolding and, crucially, this step is dynamic and continuously transitions between "folded" and "unfolded" states as opposed to predefined, unchanging states. By using a simple design it is possible to create large system sizes for a multi-lengthscale view, where the effect of a generic unfolding process on network formation can be thoroughly characterised.
\section*{Methods}
\subsection*{Model Design and Rationale}
\begin{figure}[ht!]%
\centering

    \subfloat[\centering]{{\includegraphics[width=\linewidth]{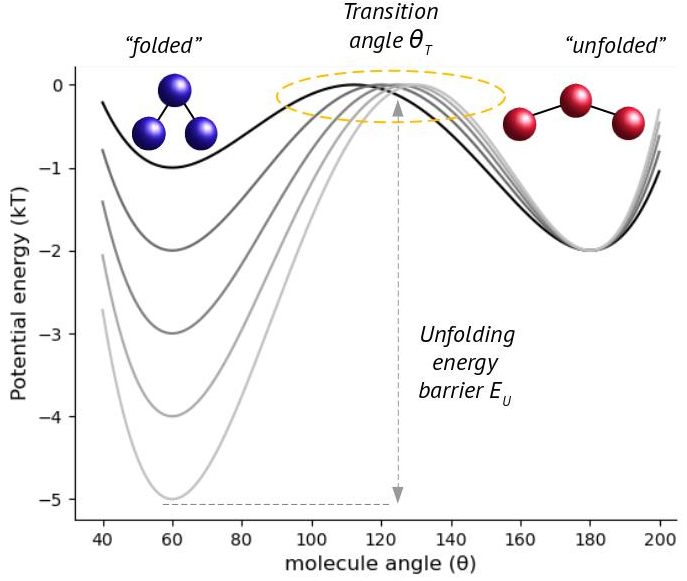} }\label{5wells}}%
    \qquad
    \subfloat[\centering]{{\includegraphics[width=\linewidth]{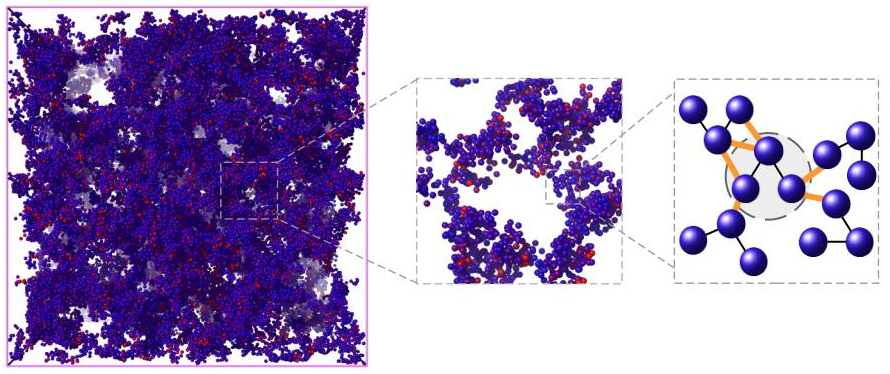} }\label{zoomzoom}}%
    \caption{(a): The double well potential for the intramolecular angle. The three-bead molecule conformation is shown for each energetic minimum. The energetic barrier of unfolding $E_U$ is varied whilst the refolding energy barrier in the reverse direction remains fixed for this work. Highlighted is the transition angle, which changes for each $E_U$. (b): A schematic of how a percolating network is formed in the simulations. A single three-bead molecule is outlined and has 6 percolating bonds highlighted in orange. Each three-bead molecule can make 6 bonds, leading to the growth of a network.}
    \label{halothreebead}%
\end{figure}

Proteins have complex 3D structures and unique multiple-step unfolding pathways. Here we use a deliberately simple model to capture basic single-step unfolding that results in shape change and geometric elongation. A three-bead molecule molecule composed of three spheres has a central particle which is permanently bonded to each of its neighbouring particles via a harmonic bond, as in Figure \ref{halothreebead}. The three-bead molecule is considered to be in the ideal `folded' state when the spheres are aligned in an equilateral shape, or when the intramolecular angle is 60\degree. The molecule is considered to be `unfolded' when the local maximum in potential energy and the transition angle $\theta$\textsubscript{T} is surpassed, as can be seen in Figure \ref{zoomzoom}. A completely unfolded molecule has an intramolecular angle of 180\degree. The molecule angle is controlled by a double well potential, with energetic minima at the "folded" and "unfolded" states. This double well potential represents the free energy landscape of a two-state protein where the states are folded and unfolded, with a transition state in-between that has a barrier height, as described by the Bell model \cite{bell1978s}. This barrier height is the energetic barrier of unfolding $E_U$ and ranges from 1 to 5kT, covering three cases where: the unfolded state is preferred; both folded and folded states are equally preferred; and the case in which the folded state is preferred, where it is more energetically favourable for the molecules to remain folded.
The range of $E_U$ has been selected to cover conditions where proteins can unfold freely at room temperature, up until a minimal number of proteins unfold. Values of $E_U$ for single proteins are often around 30-50kT\cite{hoffmann2012csr, schlierf2006bj, carrion-vazquez1999pnasu, brockwell2005bj, junker2009s}, higher than in our systems. This is due to the highly coarse-grained aspect of our protein model and the accelerated nature of coarse-grained molecular dynamics simulations, and ensures that simulations can be carried out in a feasible timescale.\\\\
Each particle in the three-bead molecule can make two intermolecular bonds, making a total of 6 possible intermolecular bonds per three-bead molecule. The reaction probability was set sufficiently high such that intermolecular bonds form instantaneously and permanently upon contact with another particle, denoting diffusion-limited aggregation (DLA)\cite{witten1981prl}. Permanent intermolecular bonds represent chemically crosslinked gelation. By using this coarse-grained, dynamic model, the computational cost for representing a large system of proteins is significantly reduced compared to other simulation techniques \cite{stradner2020sm}. There are no dihedral potentials or torsional angles to account for, allowing for more resources to be allocated to growing a large-scale network. For comparison, the previously mentioned simulations performed in BioNet were of systems of 5000 proteins\cite{hughes2021an}, whereas this model can simulate $10^4$ proteins.
\subsection*{MD Simulations}
MD simulations were performed using the simulation package LAMMPS (\hyperlink{https://www.lammps.org/}{https://www.lammps.org/})\cite{thompson2022cpc}. Thermal motion and implicit hydrodynamics were simulated by incorporating Brownian dynamics through the use of overdamped Langevin dynamics. By applying a Langevin thermostat in conjunction with velocity-Verlet time integration, the total force \textit{F} on each atom is
\begin{equation}
    F = F_c +F_f + F_r
    \label{forceseq}
\end{equation}
where \textit{F\textsubscript{c}}, \textit{F\textsubscript{f}} and \textit{F\textsubscript{r}} are the inter-particle force, the frictional force and the noise component respectively. For this work, \textit{F\textsubscript{c}} is controlled by a truncated, repulsive Lennard-Jones potential where $\sigma$ and $\epsilon$ = 1 (the simulation is run in Lennard-Jones units),
\begin{equation}
    E_r = 4\epsilon \left[\left(\frac{\sigma}{r}\right)^{12} - \left(\frac{\sigma}{r}\right)^6\right]\qquad r < r_c
    \label{ljeq}
\end{equation}
When the distance between particles $r$ is less than the cut-off distance $r_c$=2\textsuperscript{$\frac{1}{6}$}$\sigma$ ($r_c$ is also the particle radius), both particles experience a soft repulsion to prevent overlap.\\\\
A box with periodic boundary conditions was filled with three-bead protein molecules such that the volume fraction \textit{V\textsubscript{f}} of the system was 7$\%$ to be comparable to experimental systems\cite{hughes2021an}. Here the simulation box was 89$r_c$ unless otherwise stated, corresponding to 31512 three-bead molecules. This box size was selected to be sufficiently large such that interactions between high-density regions could be observed, but still small enough for reasonable simulation  and analysis times. To equilibrate the system, three-bead molecules were generated in a lattice configuration in the box and their angle potential was set to be an arbitrarily strong harmonic bond, to prevent unfolding during equilibration. After three-bead molecules were randomly distributed and randomly oriented, intermolecular bonding was then enabled and the angle potential \textit{E}\textsubscript{$\theta$} was set to Equation \ref{quarticeq}, where \textit{K\textsubscript{2}}, \textit{K\textsubscript{3}} and \textit{K\textsubscript{4}} are pre-factors and $\theta$\textsubscript{0} is the equilibrium maximum (see SI for more information)\dag. This potential represents the double well seen in Figure \ref{5wells}.
\begin{equation}
    E_\theta = K_2(\theta - \theta_0)^2 + K_3(\theta - \theta_0)^3 + K_4(\theta - \theta_0)^4
    \label{quarticeq}
\end{equation}
The simulations were run for the time required for the network to gel and relax, and were repeated for a total of 10 runs per energetic barrier of unfolding $E_U$ - the gelation time was identified as when a cluster of crosslinked proteins percolates in all three dimensions across the system boundaries. The simulations were stopped when the systems stabilised, which in this case was when the number of unfolded proteins plateaued. System states were analysed at the gelation point $T_g$ and at the end of simulation time. The percolation detection algorithm of Livraghi et al.\cite{livraghi2021jctc} was used during post-simulation analysis to find the percolation dimension of the networks across periodic boundary conditions, for each recorded frame.\\\\
To understand the effect of chemically crosslinking the network, systems with no intermolecular interactions were prepared, equilibrated and run in the same manner, only without Lennard-Jones repulsion or intermolecular bonding enabled (as detailed by Equations \ref{forceseq} and \ref{ljeq}) in order to represent pre-gel solutions. In these systems the intramolecular angle potential and the harmonic bonds were the only factors affecting protein behaviour. Over the course of all simulations the parameters tracked were: particle coordinates; three-bead intramolecular angles and angle energies; and forces of intermolecular bonded pairs.\\\\
Single molecule simulations were performed in order to convert the simulation timescale to real units. The self-diffusion constant was calculated for a single molecule and by approximating the radius of a three-bead molecule to be 30\r{A}\cite{jachimska2008l}, the simulation timescale is estimated to be around $10\mu$s. The rational behind this calculation can be found in the Supporting Information\dag. Visualisations of the system were created using PyMOL\cite{pymol}.
\section*{Results}
\begin{figure}
    \centering
    \includegraphics[width=0.9\linewidth]{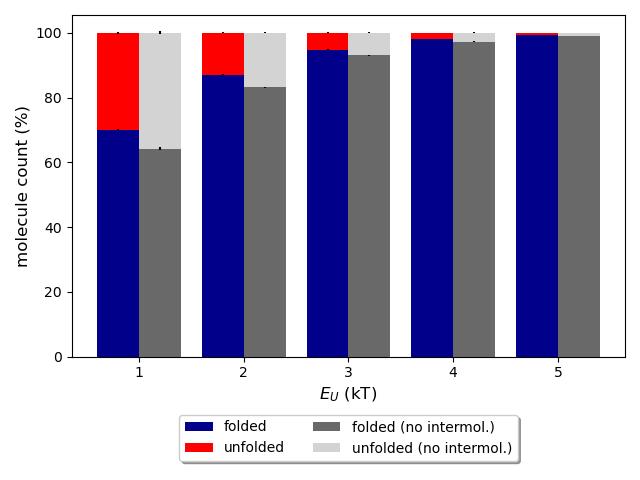}
    \caption{Unfolded vs folded populations of proteins per $E_U$ at the end of simulation. For interacting systems, "folded" proteins are coloured in blue and "unfolded" proteins are coloured in red. For systems with no intermolecular interactions enabled, "folded" proteins are coloured in light grey and "unfolded" proteins are coloured in dark grey.}
    \label{barrieranglepop}
\end{figure}

\begin{figure}
    \centering
    \includegraphics[width=\linewidth]{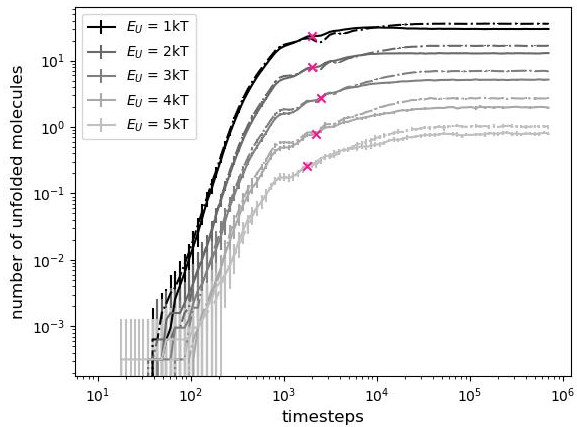}
    \caption{The total number of three-bead molecules unfolded over the course of each simulation. A molecule is considered "unfolded"  if its intramolecular angle exceeds the maximum of its angular potential. The pink crosses denote the average gelation point $T_g$ for each $E_U$. The dashed lines represent simulations without intermolecular interactions enabled.}
    \label{cumulativeunfolding}
\end{figure}

\begin{figure}[b!]
    \centering
    \includegraphics[width=\linewidth]{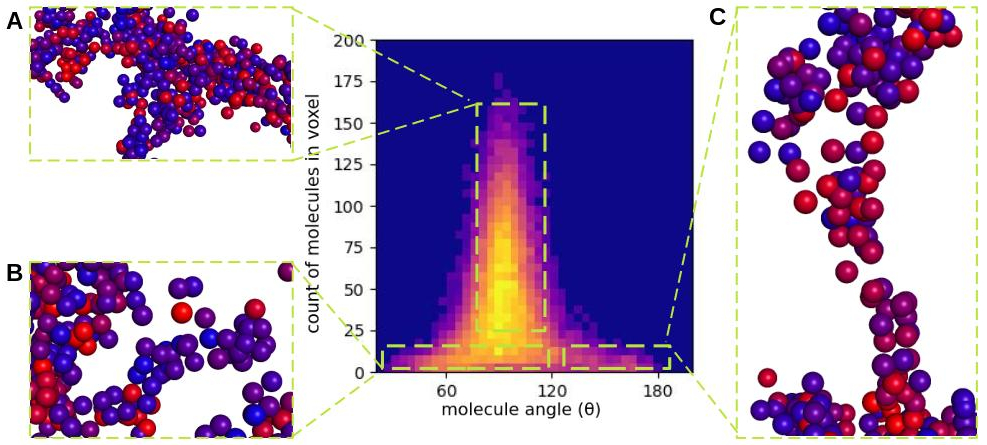}
    \caption{A heat map describing local regions of systems of chemically crosslinked protein networks with $E_U$=1kT. (a): High density regions with locally averaged "intermediate" states of unfolding, or a mixture of unfolding degrees - this is also the most common configuration for local regions in the system, as can be seen by the colouration. (b): Low density regions with folded proteins. (c): Low density regions with unfolded proteins, or strand regions.}
    \label{histogramexplained}
\end{figure}

\begin{figure*}
    \centering
    \includegraphics[width=\textwidth]{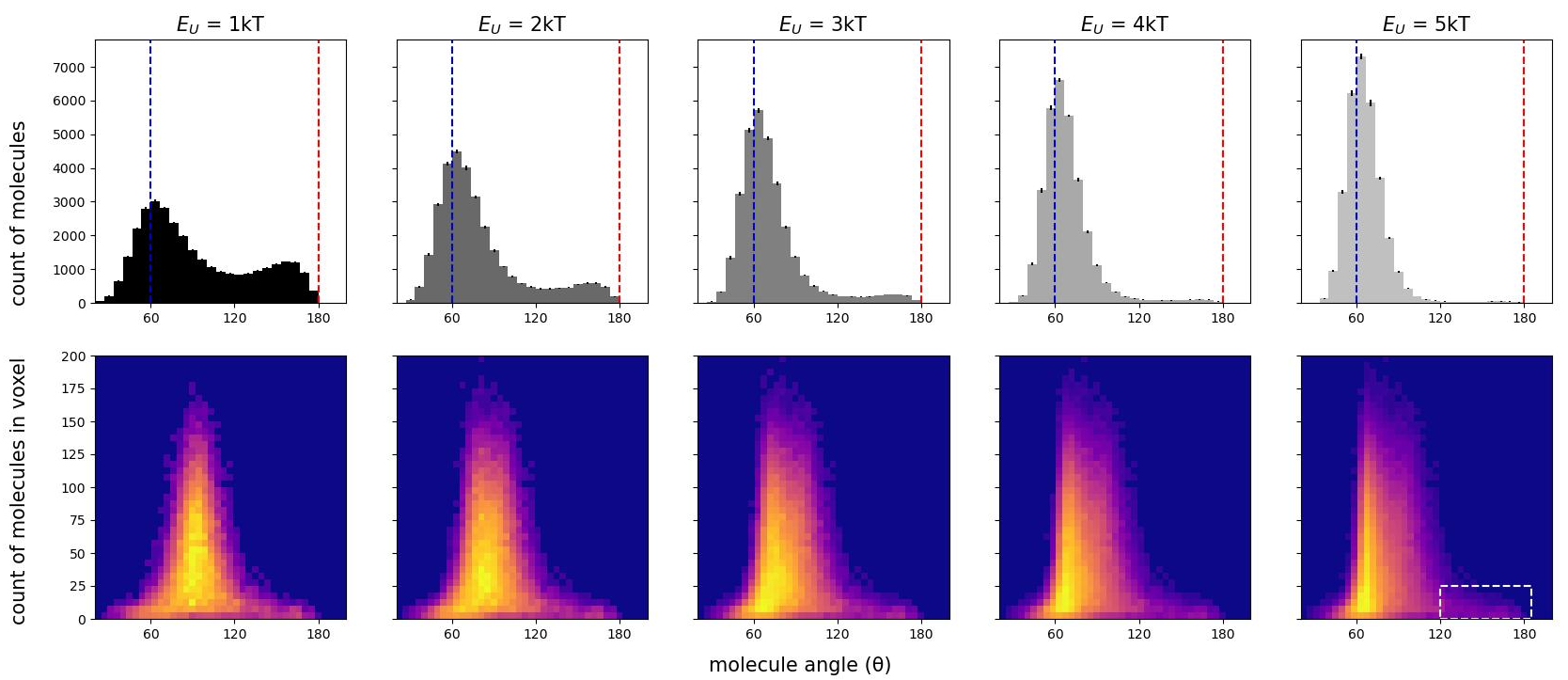}\
    \caption{Top: Histogram of average intramolecular angle distribution per energetic barrier of unfolding $E_U$ at the end of simulation. The dashed vertical lines in blue and red indicate the equilibrium angles for folded and unfolded proteins respectively. Bottom: Heat maps describing average degree of unfolding and local population count for each $E_U$, plotted by dividing the simulation box into voxels. Regions in yellow indicate the most popular local structure of a voxel. An example of a low density, highly unfolded strand region is in the white dashed box in the heat map for $E_U$=5kT. Top and bottom figures are not directly comparable as the top depicts the mode angle of individual molecules, whereas the bottom depicts the mean angles averaged over small regions of space.}
    \label{anglehist}
\end{figure*}

We investigated the effect of systematically varying the energetic barrier of unfolding $E_U$ and how this parameter impacts the formation of a crosslinked protein network structure over time.
\begin{figure*}[ht]
    \centering
    \includegraphics[width=\textwidth]{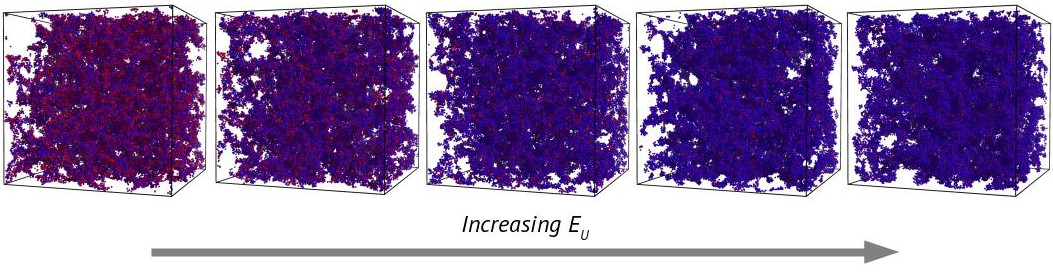}
    \caption{Snapshots of interacting systems at the end of simulations with increasing energetic barrier of unfolding $E_U$. "Folded" proteins are represented by blue spheres. "Unfolded" proteins are represented by red spheres. On the left is a system with $E_U$=1kT and on the right is a system with $E_U$=5kT.}
    \label{increasingeu}
\end{figure*}
\begin{figure}[h]%
    \centering
    \subfloat[\centering]{{\includegraphics[width=0.45\linewidth]{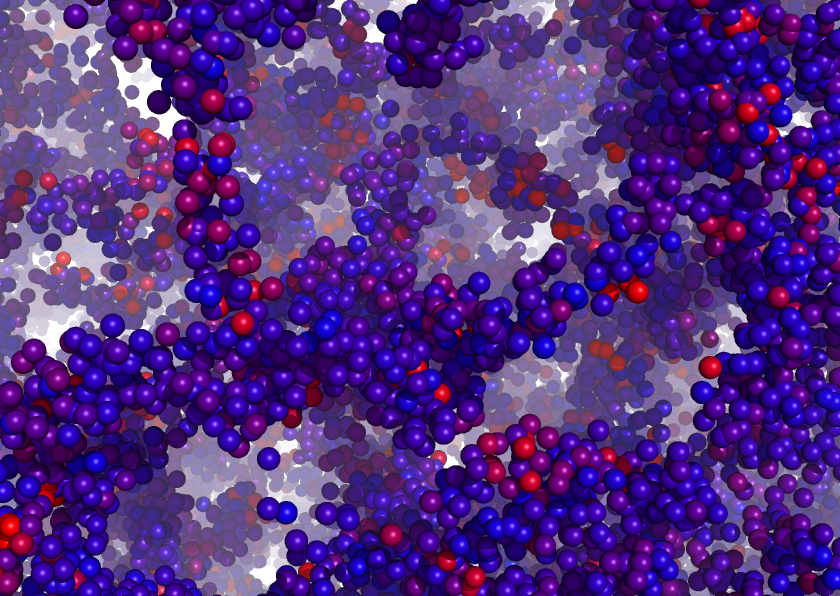} }\label{heterogeneity}}%
    \qquad
    \subfloat[\centering]{{\includegraphics[width=0.45\linewidth]{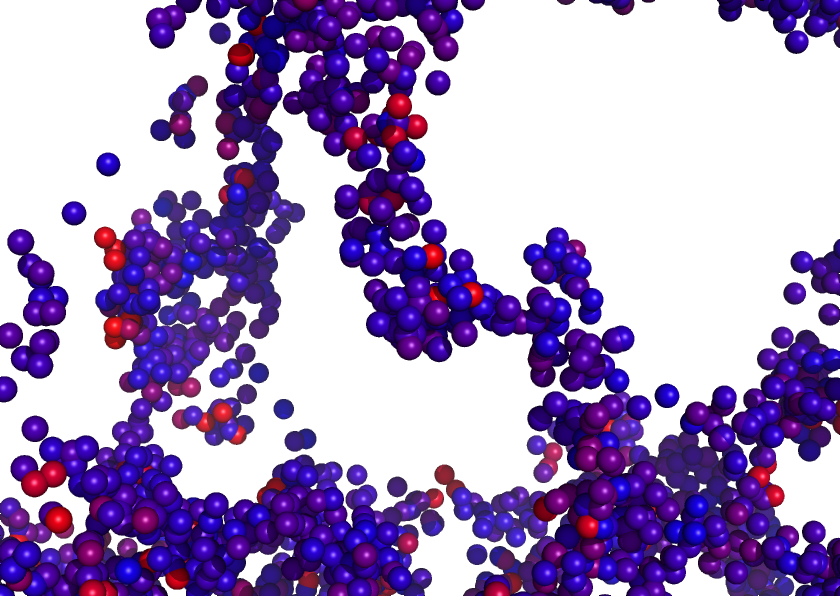} }\label{sortofstrand}}%
    \caption{Snapshots of a formed interacting network where $E_U$=3kT. "Folded" proteins are represented by blue spheres. "Unfolded" proteins are represented by red spheres. (a): A zoomed-in image of the completely formed interacting network showing depth, structural heterogeneity and variation in unfolding degree. (b): An example of a low density region, taken from a slice of the network for visual purposes. Links between beads are not explicitly shown.}
    \label{visexample}
\end{figure}
All systems with intermolecular interactions produced heterogeneous, crosslinked networks for all measured $E_U$ as seen in Figure \ref{increasingeu}. Further visual inspection of the simulations show heterogenous networks composed of high-density regions of proteins (clusters) with low-density regions of proteins (inter-cluster regions) connecting these groups together. Both folded and unfolded proteins coexist, with the inter-cluster regions appearing visually to have a higher fraction of unfolded, rather than folded, proteins in the system. Figure \ref{visexample} shows examples of these regions.
\paragraph*{Fewer proteins unfold with increasing $E_U$.} Figure \ref{barrieranglepop} shows the proportion of folded to unfolded molecules for each $E_U$ at the final frame of the simulation. It can be seen that as $E_U$ increases the quantity of unfolded proteins decreases, to the point where at $E_U$=5kT there are only 0.79$\pm$0.04$\%$ of unfolded molecules. Error is measured as standard deviation over the 10 runs per $E_U$. Simulations were also performed without intermolecular interactions where the only energetic contributions arose from intramolecular angle potential and harmonic intramolecular bonds. Again as $E_U$ increases the quantity of unfolded proteins decreases. However in the case of interacting systems, more proteins are in the folded state than in non-interacting systems - evidencing local crowding effects in high density regions \cite{wang2012jacs, das2024ijobm, olgenblum2026ps}. These effects emerge as a result of high-density regions in which proteins may experience different intermolecular interactions to those in low-density regions. This is further shown by changes to the cumulative fraction of unfolded proteins, which will now be discussed.
\paragraph*{Intermolecular interactions control network rearrangements.} Figure \ref{cumulativeunfolding} shows the cumulative number of unfolded proteins during the simulations for each $E_U$. Over the course of the simulations the number of unfolded proteins increases non-monotonically. Fewer proteins unfolded with increasing $E_U$, as previously discussed and in accord with intuition. We additionally observe that for $E_U$=3, 4, 5kT (where the "folded" state is preferred as in Figure \ref{5wells}), the gelation time decreases with increasing $E_U$.\\\\
It can be seen in Figure \ref{cumulativeunfolding} that for each $E_U$ there are fewer unfolded proteins in systems with intermolecular interactions enabled - this occurs mostly after gelation and  during network rearrangement. The non-monotonic shape of the curve shows that some protein refolding occurs over the course of the simulation in both interacting and non-interacting systems - this effect is further visible in Figure S1\dag. For non-interacting systems (most visible in $E_U$=1kT and 2kT), some refolding can be seen occurring just after what would be the gelation point $T_g$. $T_g$ is a pivotal moment in network behaviour: up until gelation, the number of unfolded proteins grows exponentially. At time $> T_g$, there is a divergence where non-interacting systems have a larger proportion of unfolded proteins relative to the interacting systems. In interacting systems, proteins that are locally crowded are bound to their position due to intermolecular bonds and can only rearrange by refolding, rather than moving. In non-interacting systems proteins are not locally bound, allowing them to refold in a homogenous space and minimize the system's free energy - this refolding allows for further protein unfolding as the system stabilises. The energetic barrier for refolding is roughly the same for all systems (see Figure \ref{5wells}), meaning that by varying $E_U$ there is control over the competition between unfolding and refolding.
\begin{figure*}[ht]%
\centering
    \subfloat[\centering]{{\includegraphics[width=0.47\linewidth]{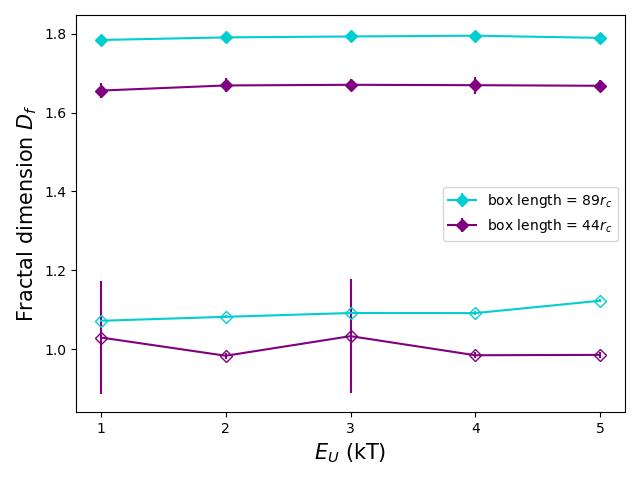} }\label{fracdim}}%
    \qquad
    \subfloat[\centering]{{\includegraphics[width=0.47\linewidth]{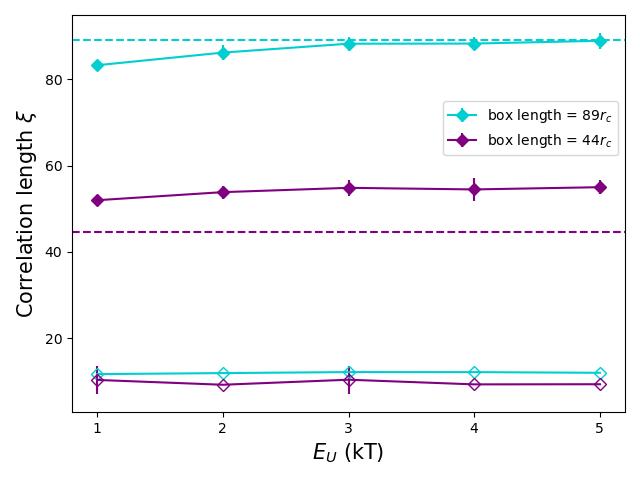} }\label{corrlength}}%
    \caption{Average fractal dimension \textit{D\textsubscript{f}} and correlation length $\xi$ of chemically crosslinked interacting protein networks at each energetic barrier of unfolding $E_U$. In blue are systems of box length = 89$r_c$ and in purple are systems of box length = 44$r_c$. Open symbols denote values at $T_g$ and closed systems denote values at the end of the simulation. (a): Fractal dimension \textit{D\textsubscript{f}} for each system size and $E_U$. (b): The correlation length $\xi$ for each system size and $E_U$, which is shown in reference to particle radii $r_c$. The system size is shown with a dashed line.}
    \label{corrlengthfracdim}
\end{figure*}
\\\\Additional metrics to aid the understanding of the network's behaviour during formation and relaxation are the fractal dimension $D_f$ and the correlation length $\xi$. In this study the fractal dimension is used as a parameter to describe how complex (for example, ramified or compact) the system structure is. $D_{f}$ is calculated by using a standard box counting algorithm\cite{hagiwara1998}. This algorithm calculates the minimum number of unique boxes required to cover the network and uses both this number and the size of the boxes to calculate $D_f$. The correlation length $\xi$, which we interpret as a characteristic cluster size, is calculated by finding the inflection point of fits to the plots produced using the box counting algorithm. An example of a plot and its fit can be seen in Figure S2\dag. Figure \ref{corrlengthfracdim} shows both of these quantities at the point of gelation - and at the end of simulation time - for two different system sizes, namely with box edge lengths of 89$r_c$ (as used for the data above) and 44$r_c$. For both system sizes it can be seen that $D_{f}$ and $\xi$ increase post-gelation. At gelation, both systems exhibit $D_{f}\sim1$, which is the lowest for which percolation can be expected and suggests the system contains a ramified, stranded system-spanning cluster at gelation.\\\\
Although $D_{f}$ slightly decreases with increasing $E_U$, this variation is weak. In this system $\xi$ increases after network gelation and the subsequent relaxation, and slightly decreases with $E_U$. It should be noted that when calculating $\xi$ for various systems its magnitude may exceed the box size; this is because $\xi$ is calculated by fitting to the box counting data and is not strictly bounded by the box dimensions. In our simulations with these metrics to describe the structure, clusters increase in size with increasing $E_U$ but have consistent complexity ($D_{f}$) across varying $E_U$\cite{hughes2025jocais}. The largest factor influencing structure is not $E_U$ but time; that is, is the process of relaxation, in which the three-bead molecules can rearrange to reduce their free energy. As each system relaxes after $T_g$, there is a significant increase in both $D_{f}$ and $\xi$ due to local rearrangements, allowing clusters to both grow in size and density \cite{vanvliet2000h}.
\paragraph*{Proteins have a broader distribution of unfolding degree at low $E_U$.} Figure \ref{anglehist} shows the distribution of intramolecular angles in a network at the end of the simulation. For $E_U$=1kT there is a bimodal distribution of intramolecular angles with peaks around 60\degree and 160\degree. The peak around 160\degree denotes the mode unfolded intramolecular angle, whereas 180\degree is the unfolded intramolecular angle that is most energetically favourable for a single three-bead molecule. As $E_U$ increases, the peak around 60\degree grows in height whilst the peak at 160\degree shrinks - the bimodal distribution becomes more unimodal and the proportion of proteins with an intermediate intramolecular angle decreases. To gain further understanding of the distribution of unfolded proteins in varying density regions, heat maps of density and internal bond angle averaged within local regions were produced to further characterise the structure - these heat maps provide quantitative comparison of occurrences of each local configuration. Figure \ref{histogramexplained} gives examples of local configurations corresponding to features in the heat maps to aid interpretation, and the heat maps themselves for systems with increasing $E_U$ can be seen in Figure \ref{anglehist}.
Precisely, the simulation boxes were divided into voxels with side length 6$r_c$, meaning that two unfolded molecules can span each edge of the voxel. The average intramolecular angle of molecules within the voxel were calculated alongside the count of molecules within that voxel. At $E_U$=1kT the voxel molecule count is approximately normally distributed against the intramolecular angle, indicating that the central limit theorem applies and the numbers of folded and unfolded molecules in each voxel are not strongly correlated, giving rise to the Gaussian shape. As $E_U$ increases, the distribution becomes more skewed and the average intramolecular angle decreases, suggesting that  though there is a smaller unfolded population, systems with higher $E_U$ deviate less from the equilibrium folded or unfolded states. This results in two intramolecular angle means that cause this deviation from the Gaussian form, resulting from unfolding in the network being spacially correlated.\\\\
Furthermore, Figure \ref{anglehist} also highlights a tail of unfolded proteins in low density regions that persists as $E_U$ increases. The existence of this tail confirms the earlier observation that the low-density inter-cluster regions have an enhanced population of unfolded molecules; this was discussed in the context of Figure \ref{visexample}. Although these unfolded strand regions become less common as $E_U$ increases, they are still a key network feature visible in all systems and in all heat maps. As such, this is an indication that unfolding occurs consistently in low-density regions connecting high-density regions whenever a system has proteins that can unfold.
\section*{Discussion}
Figure \ref{corrlengthfracdim} indicates that hydrogel structure is governed primarily by post-gelation relaxation, whereas $E_U$ exerts a relatively minor influence. Progressing from gelation to the gel network at the end of simulation, we observe an approximately nine-fold increase in $\xi$, and $D_{f}$ increases by a factor of roughly 1.5. The same trend is also observed in smaller systems with a box length of 44$r_c$. Following relaxation, the correlation length $\xi$ approaches the size of the simulation box and the fractal dimension $D_f$ increases. However, $D_{f}$ remains lower overall than in the larger system. This suggests that the rearrangements act to increase $D_{f}$ monotonically in time, with the final $D_{f}$ being observed when $\xi$ reaches the box dimensions and the relaxation processes are arrested. For a smaller system size this arrest should happen sooner, resulting in the observed lower $D_f$ than in a larger system size. For such a simple system, this large structural rearrangement is unexpected. In a previous lattice-based model there was only a minimal increase in $D_{f}$ between percolation and the final gel state\cite{cook2023sm}. As the current model is off-lattice, the observed rearrangement effect may be due to propagation of small perturbations and rearrangements throughout the network.\\\\
To compare this work to experimental studies of crosslinked protein hydrogels, BSA shall be used as a model protein, BSA is a globular protein commonly used as a model for protein hydrogels\cite{boye, hagiwara1998, khoury, nikfarjam2023b, olgenblum2026ps, wu2001l} and can be described as having 3 ellipsoidal / spherical regions \cite{kuhar2020ac}. In previous work, the effects of unfolding during network formation were studied in BSA gels\cite{hughes2021an}. For \textit{in-vitro} gels, proteins with the capacity to unfold in pre-gel solutions did not undergo unfolding due to their thermodynamic stability. This contrasts the findings in these simulations, where not only do proteins unfold in non-interacting systems but more of them unfold than in interacting, percolated systems. However 28$\%$ of proteins unfolded in gels with an increased probability of single protein unfolding, comparable to 30.0$\pm$0.3$\%$in our simulated system. In both interacting systems and BSA gels with an increased proability of unfolding, $\xi$ marginally increases but in simulations there is a higher contrast between $\xi$ at $T_g$ and post-relaxation - there is an even greater difference in \textit{D\textsubscript{f}} between \textit{in-vitro} and \textit{in-silico} systems. For BSA gels with an increased probability of unfolding, the number of folded proteins decays over time without any refolding occurring, and with minimal folded protein decay for BSA gels that have not been solvated. The differences between previous and current work suggest that this simple model is insufficient to fully represent factors inducing and describing protein unfolding in a gel. Though this three-bead molecule model involves a dynamic unfolding step, it does not accurately describe the complex multi-step unfolding pathway of a protein such as BSA\cite{kuhar2020ac}. Such a small geometric change due to many fewer degrees of freedom means that the entropic difference in the folded and unfolded states in this model is much lower and practically negligible in comparison to that of gels formed \textit{in-vitro}. Additionally, water is modelled implicitly in this work and so negates how osmotic pressure may affect the unfolding pathway of the protein. Despite this, we are still able to capture the heterogeneous protein network structure which has been observed experimentally and witness the how the first stages of unfolding can affect network structure and behaviour in protein hydrogels.
\section*{Conclusions}
In this work we have developed a coarse-grained protein model that captures a heterogenous chemically crosslinked protein network formed from proteins that are able to dynamically fold and refold. This model provides insight into network formation and relaxation. By introducing dynamic conformational change it is possible to control the extent of protein unfolding during network formation. Whilst in this model network structure is not strongly affected by the energetic barrier of unfolding, it is instead affected by post-gelation relaxation. Protein refolding has also been observed in interacting systems, where it may be driven by local crowding effects. By contrast, in non-interacting systems refolding occurs earlier and allows proteins to undergo further unfolding as the systems relaxes and reorganises. Analysis of protein intramolecular angles evidenced that in systems with proteins that can unfold, unfolding did occur in low density regions but also occurred to a lesser extent in medium density regions. Systems with more mechanically robust proteins preferred to have more tightly folded proteins, yet still had local low density populations of unfolded proteins - we anticipate that these connective stranded low-density regions between high density regions will hold a greater role when the system experiences mechanical perturbations. This study provides a framework to understand from first principles the role of unfolding within protein networks, and lays the foundation of how a protein hydrogel's mechanical response may be impacted by dynamic unfolding.

\section*{Author contributions}
V.B - Methodology, software, data curation, formal analysis, investigation, validation, visualisation, writing - original draft, writing - review and editing; L.D: conceptualisation, investigation, writing - review and editing, project administration, supervision; D.H: conceptualisation, methodology, software, investigation, resources, validation, writing - review and editing, project administration, supervision.

\section*{Conflicts of interest}
There are no conflicts to declare.

\section*{Data availability}

Input files and raw data produced for this work are available at the University of Leeds data repository \hyperlink{https://doi.org/10.5518/1927}{https://doi.org/10.5518/1927} (subject to review). 

\section*{Acknowledgements}
V.B is supported by an Engineering and Physical Sciences Research Council (EPSRC) PhD studentship through the Soft Matter for Formulation and Industrial Innovation Centre for Doctoral Training (SOFI$^2$ CDT) (EP/S023631/1). L.D is supported by a European Research Council Consolidator Fellowship / UKRI Frontier Research Fellowship for the MESONET project (EUKRI EP/X023524/1). This work was carried out using the Aire High-Performance Computing system at the University of Leeds, UK. We are grateful to the Dougan group for helpful discussions, particularly Dr Matt Hughes for the helpful input and discussion in the interpretation of the cumulative unfolding data and the local structure heat maps.


\balance


\bibliography{allrefs} 
\bibliographystyle{rsc} 

\end{document}


\title{\textbf{Mesoscale heterogeneity in protein hydrogels induced by dynamic unfolding and post-gelation rearrangements}\\Supporting Information}
\author{Victoria Byelova, Lorna Dougan, David Head}
\date{}
\maketitle
\begin{figure}
    \centering
    \includegraphics[width=\linewidth]{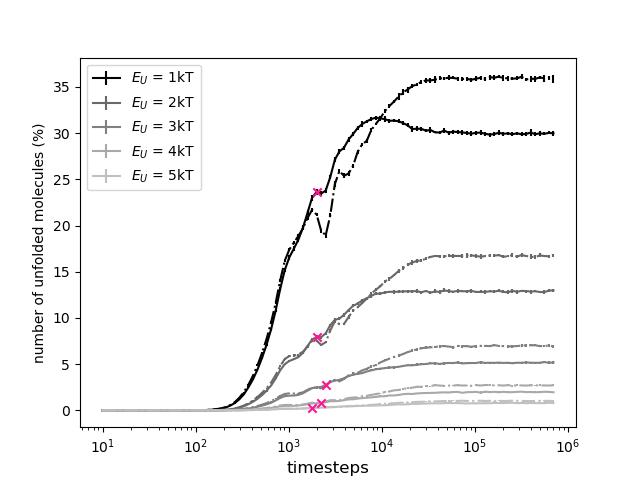}
    \caption{A plot showing the count of cumulative unfolded proteins over the course of the simulation run time. The axis is semi-logarithmic and more clearly shows the refolding behaviour in the systems.}
    \label{semilog}
\end{figure}
\begin{figure}
    \centering
    \includegraphics[width=\linewidth]{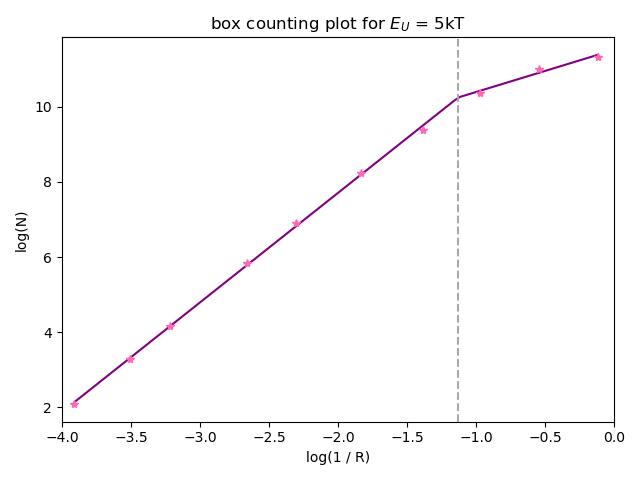}
    \caption{An example box counting plot produced for finding the fractal dimension $D_f$ and the correlation length $\xi$. This plot is for $E_U$=5kT. A dotted line indicates where the system lengthscale changes and is where $\xi$ is acquired. \textit{R} is the size of the box used to cover the system, \textit{N} is the number of unique boxes required to cover the system.}
    \label{boxcounting}
\end{figure}
\subsection*{Real time conversion}
In order to calculate an estimate for the real time of the simulation, single three-bead molecule simulations were performed. The energetic barrier of unfolding $E_U$ was set to 100kT and unwrapped coordinates were tracked for a sufficiently long time, such that the molecule could freely diffuse its own length multiple times. The mean-squared deviation was calculated and averaged over 20 runs. Following this, the simulation self-diffusion constant was converted into real time with an approximation of 30\r{A} for the molecule radius, 0.89 x 10$^{-3}$ \textit{Pa s} for the water viscosity at 298\textit{K}.